\documentclass{ifacconf}

\usepackage{graphicx}      % include this line if your document contains figures
\usepackage{natbib}        % required for bibliography
\usepackage{amssymb}
\usepackage{lipsum}
\usepackage{amsmath}
\usepackage{gensymb}
\usepackage{float}
\usepackage{graphicx}
\usepackage{subcaption}
\usepackage{tabularx}
\usepackage{comment}
\usepackage{dblfloatfix}
\usepackage[table,xcdraw]{xcolor}
\usepackage{wrapfig}
\usepackage{soul}
\usepackage{multirow}
\usepackage{cuted}
\usepackage{tikz}

\begin{document}
\begin{frontmatter}

\title{Decentralized Motion and Resonant Damping Control for High-Bandwidth and Cross-Coupling Reduction in MIMO Nanopositioners} 

\thanks[footnoteinfo]{This work was financed by Physik Instrumente (PI) SE \& Co. KG and co-financed by Holland High Tech with PPS Project supplement for research and development in the field of High Tech Systems and Materials.}

\author[First]{Aditya Natu} 
\author[First]{Hassan HosseinNia} 

\address[First]{Department of Precision and Microsystems Engineering, Delft University of Technology, Mekelweg 2, 2628 CD Delft,
The Netherlands (e-mail:  a.m.natu@tudelft.nl; s.h.hosseinniakani@tudelft.nl)}

\begin{abstract}                % Abstract of not more than 250 words.
Piezoelectric nanopositioning systems are widely used in precision applications that require nanometer accuracy and high-speed motion; however, lightly damped resonances and pronounced cross-axis coupling severely limit bandwidth and disturbance rejection. This paper presents a decentralized dual-loop control strategy for a two-axis nanopositioner, combining an inner non-minimum-phase resonant damping controller with an outer motion controller on each axis. The dominant diagonal resonance is actively damped to enable closed-loop bandwidths beyond the first structural mode, while a parallel band-pass damping path is specifically tuned to a higher-order resonance that predominantly affects the cross-coupling channels. Experimental results demonstrate that this targeted band-pass damping substantially reduces cross-axis coupling and enhances disturbance rejection, without compromising tracking accuracy.
\end{abstract}


\end{frontmatter}
%===============================================================================

% Color Commands
% --- MATLAB Color Definitions ---
\definecolor{matlabblack}{rgb}{0,0,0}
\definecolor{matlaborange}{rgb}{0.8500,0.3250,0.0980}
\definecolor{matlabgreen}{rgb}{0.4660,0.6740,0.1880}
\definecolor{matlabmaroon}{rgb}{0.6350,0.0780,0.1840}
\definecolor{matlabyellow}{rgb}{0.9290,0.6940,0.1250}
\definecolor{matlabblue}{rgb}{0,0.4470,0.7410}
\definecolor{matlabpurple}{rgb}{0.4940,0.1840,0.5560}

% --- MATLAB Black ---
\newcommand{\blackline}{\raisebox{2pt}{\tikz{\draw[-,matlabblack,solid,line width=1.5pt](0,0)--(3mm,0);}}}
\newcommand{\blacklinedashed}{\raisebox{2pt}{\tikz{\draw[-,matlabblack,dashed,line width=1pt](0,0)--(3mm,0);}}}
\newcommand{\blacklinedotted}{\raisebox{2pt}{\tikz{\draw[-,matlabblack,dotted,line width=1pt](0,0)--(3mm,0);}}}

% --- MATLAB Orange ---
\newcommand{\orangeline}{\raisebox{2pt}{\tikz{\draw[-,matlaborange,solid,line width=1.5pt](0,0)--(3mm,0);}}}
\newcommand{\orangelinedashed}{\raisebox{2pt}{\tikz{\draw[-,matlaborange,dashed,line width=1pt](0,0)--(3mm,0);}}}
\newcommand{\orangelinedotted}{\raisebox{2pt}{\tikz{\draw[-,matlaborange,dotted,line width=1pt](0,0)--(3mm,0);}}}

% --- MATLAB Green ---
\newcommand{\greenline}{\raisebox{2pt}{\tikz{\draw[-,matlabgreen,solid,line width=1.5pt](0,0)--(3mm,0);}}}
\newcommand{\greenlinedashed}{\raisebox{2pt}{\tikz{\draw[-,matlabgreen,dashed,line width=1pt](0,0)--(3mm,0);}}}
\newcommand{\greenlinedotted}{\raisebox{2pt}{\tikz{\draw[-,matlabgreen,dotted,line width=1pt](0,0)--(3mm,0);}}}

% --- MATLAB Maroon ---
\newcommand{\maroonline}{\raisebox{2pt}{\tikz{\draw[-,matlabmaroon,solid,line width=1.5pt](0,0)--(3mm,0);}}}
\newcommand{\maroonlinedashed}{\raisebox{2pt}{\tikz{\draw[-,matlabmaroon,dashed,line width=1pt](0,0)--(3mm,0);}}}
\newcommand{\maroonlinedotted}{\raisebox{2pt}{\tikz{\draw[-,matlabmaroon,dotted,line width=1pt](0,0)--(3mm,0);}}}

% --- MATLAB Yellow ---
\newcommand{\yellowline}{\raisebox{2pt}{\tikz{\draw[-,matlabyellow,solid,line width=1.5pt](0,0)--(3mm,0);}}}
\newcommand{\yellowlinedashed}{\raisebox{2pt}{\tikz{\draw[-,matlabyellow,dashed,line width=1pt](0,0)--(3mm,0);}}}
\newcommand{\yellowlinedotted}{\raisebox{2pt}{\tikz{\draw[-,matlabyellow,dotted,line width=1.5pt](0,0)--(3mm,0);}}}

% --- MATLAB Blue ---
\newcommand{\blueline}{\raisebox{2pt}{\tikz{\draw[-,matlabblue,solid,line width=1.5pt](0,0)--(3mm,0);}}}
\newcommand{\bluelinedashed}{\raisebox{2pt}{\tikz{\draw[-,matlabblue,dashed,line width=1pt](0,0)--(3mm,0);}}}
\newcommand{\bluelinedotted}{\raisebox{2pt}{\tikz{\draw[-,matlabblue,dotted,line width=1pt](0,0)--(3mm,0);}}}

% --- MATLAB Purple ---
\newcommand{\purpleline}{\raisebox{2pt}{\tikz{\draw[-,matlabpurple,solid,line width=1.5pt](0,0)--(3mm,0);}}}
\newcommand{\purplelinedashed}{\raisebox{2pt}{\tikz{\draw[-,matlabpurple,dashed,line width=1pt](0,0)--(3mm,0);}}}
\newcommand{\purplelinedotted}{\raisebox{2pt}{\tikz{\draw[-,matlabpurple,dotted,line width=1pt](0,0)--(3mm,0);}}}

\section{Introduction}

Piezoelectric-actuated nanopositioning systems are central to applications such as atomic force microscopy (AFM), precision manufacturing, optical alignment, and biomedical manipulation, where nanometer-scale accuracy and high-speed motion are required (\cite{fleming2014design}). These systems typically exhibit lightly damped structural resonances, which fundamentally limit the closed-loop bandwidth and disturbance rejection (\cite{devasia2007survey}). As the control bandwidth approaches the first resonance, the associated phase lag and gain peaking can severely constrain tracking performance and may lead to closed-loop instability (\cite{sebastian2005design}).

To overcome these limitations, a variety of active damping strategies have been developed to suppress lightly damped resonant modes. Common approaches include Positive Position Feedback (PPF) (\cite{fanson1990positive}), Integral Resonant Control (IRC) (\cite{bhikkaji2008integral}), Resonant Control (RC) (\cite{ling2019robust}), and the non-minimum-phase resonant controller (NRC) (\cite{natu2024non}). These controllers introduce targeted damping around structural resonances and, when embedded in dual-loop architectures—where an inner damping loop increases the effective structural damping and an outer motion-control loop is used to achieve high tracking bandwidth—have been shown to substantially improve nanopositioning performance (\cite{eielsen2013damping}).

However, extending these methodologies to multi-axis nanopositioning platforms introduces additional challenges. Cross-coupling between axes—arising from mechanical compliance and structural interactions—causes reference or disturbance inputs in one axis to induce motion in the others, degrading point-to-point accuracy and raster-scanning precision (\cite{al2021modeling}). Effective control of multiple-input multiple-output (MIMO) nanopositioners, therefore, requires not only high performance in the diagonal channels but also the mitigation of disturbance propagation and dynamic coupling effects.

Classical methods to mitigate cross-coupling in multi-axis nanopositioners include decoupling matrices, disturbance observers, and centralized multivariable control (\cite{liu2019review,huang2024design,mohamed2014feedforward}). In decoupling or inversion-based schemes, the off-diagonal terms are compensated using an approximate inverse or pre-compensator, so that each input predominantly excites a single output. Although such approaches can substantially reduce cross-axis interactions, they depend on the accurate identification of multiple resonant modes and delays in all channels. As the number of modes and degrees of freedom increases, the resulting centralized controllers become high-order and computationally demanding, and their robustness degrades in the presence of modeling errors and unmodeled higher-order dynamics.

To reduce this complexity, recent work has explored control architectures that integrate motion and damping controllers while retaining relatively simple implementations (\cite{das2013mimo,das2015multivariable,ling2018integrating}). However, most of these studies primarily target resonances visible in the diagonal transfer functions, effectively assuming that the dominant diagonal modes also govern cross-coupling behavior. In practice, higher-order resonant modes—though not dominant in the diagonal transfer functions due to system symmetry—can strongly influence cross-coupling dynamics, facilitating disturbance propagation between axes and degrading control performance, even when the primary resonances are well damped.

In this work, a decentralized motion and resonant damping control framework is developed for a dual-loop MIMO nanopositioning system to simultaneously achieve high-bandwidth motion control and cross-coupling reduction. Each axis employs an inner-loop non-minimum-phase resonant damping controller to suppress its dominant resonance, enabling an outer motion controller to attain bandwidths beyond the first structural mode. To further mitigate disturbance propagation mediated by higher-order modes, the damping structure is extended with a parallel band-pass damping path that specifically targets a second resonance that appears predominantly in the cross-coupling channels.

Experimental validation on a two-axis piezoelectric nanopositioner demonstrates that the proposed decentralized design substantially improves closed-loop damping and cross-axis isolation. Frequency-domain sensitivity measurements show reduced cross-coupling near the targeted resonances, with only a modest change in closed-loop bandwidth, while time-domain experiments confirm improved disturbance rejection and preserved tracking accuracy for raster-like reference trajectories.

The remainder of this paper is organized as follows: Section~\ref{sec:System Description and Identification} describes the nanopositioning system and its frequency response identification. Section~\ref{sec:DesignMethodology} details the proposed control design methodology, including the decentralized damping and motion controller structures. Section~\ref{sec:ExperimentalResults} presents experimental results in both the frequency and time domains, highlighting the achieved performance improvements. Finally, Section~\ref{sec:Conclusion} concludes the paper.

\section{System Description and Identification}
\label{sec:System Description and Identification}
\begin{figure}[t!]
    \centering
    \includegraphics[width =1\linewidth]{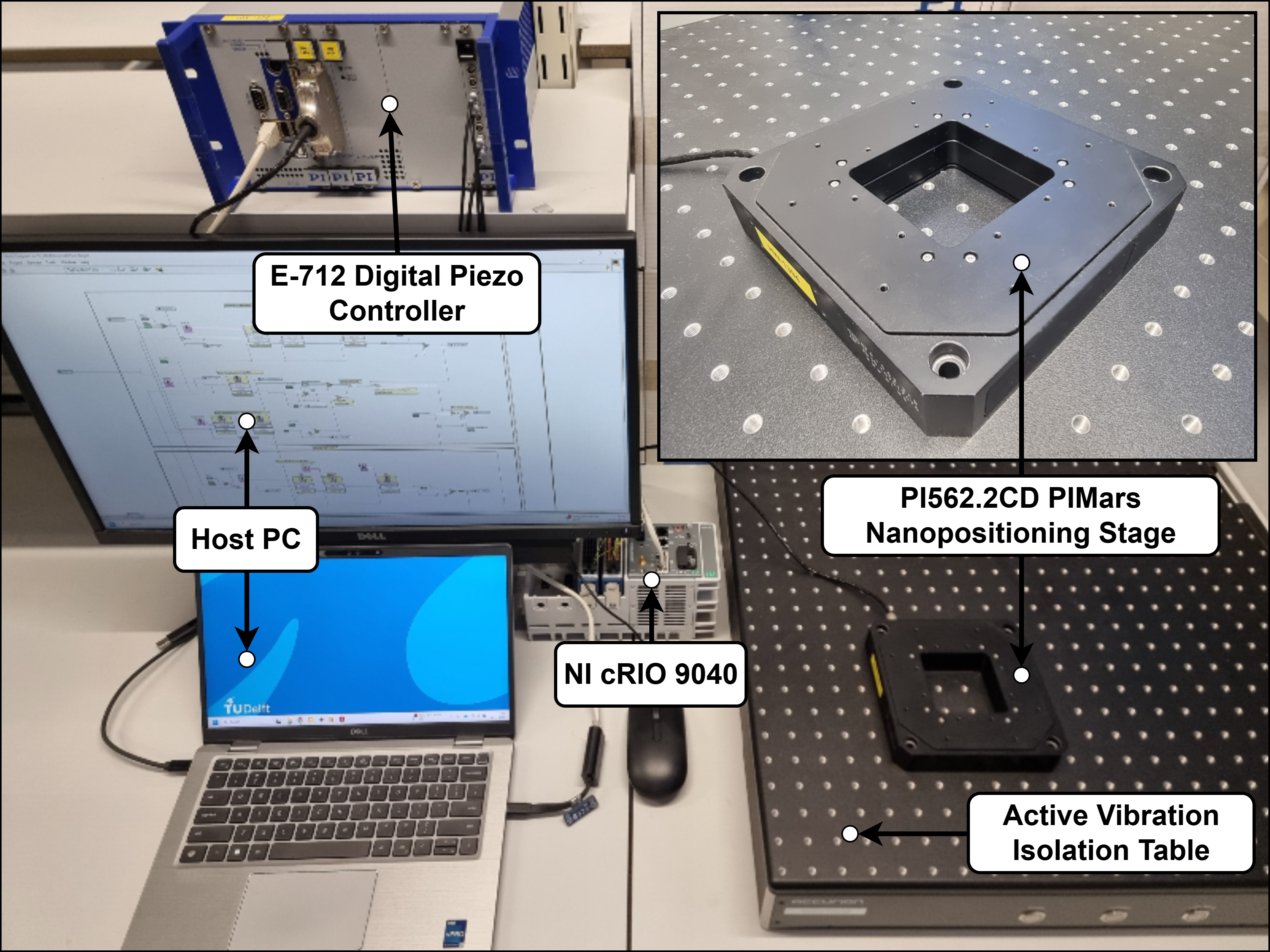}
    \caption{Experimental setup with P-562.2CD PIMars nanopositioning stage.}
    \label{fig:ExperimentalSetup}
\end{figure}

The experimental setup in Fig.~\ref{fig:ExperimentalSetup} uses a commercial P-562.2CD PIMars nanopositioning stage with a 200~$\mu$m travel range per axis. Each axis incorporates a ceramic-insulated multilayer piezo-stack actuator, a flexure-guided parallel-kinematics mechanism for frictionless motion, and high-resolution capacitive sensors for displacement feedback. The stage is driven through a voltage amplifier and signal-conditioning modules integrated into the E-712 piezo-controller. External control is provided via an NI CompactRIO chassis equipped with an embedded FPGA for real-time actuation and 16-bit analog I/O for signal transmission and acquisition. The control algorithm is implemented in LabVIEW and executed from the host computer. The actuation range is 0–10~V, with a sampling time $t_s = 30~\mu$s ($f_s = 33.\overline{33}$~kHz, $f_N = 16.\overline{66}$~kHz), which is well above the frequency range of interest (1–1000~Hz).

A sinusoidal chirp signal (0.75–0.8~V) was generated in LabVIEW and applied to the piezo-actuators for system identification. The capacitive sensor outputs were recorded and processed in MATLAB, where frequency response functions (FRFs) were estimated using a Hanning window to reduce spectral leakage. The high sampling rate and strong coherence enabled accurate identification over 1–1000~Hz. The dominant resonance appears near 165~Hz in each axis, while the second mode at 231~Hz is more pronounced in the cross-coupling terms, as shown in Fig.~\ref{fig:ClosedLoopFRF}. The phase lag below the first resonance is due to actuator–amplifier dynamics and inherent delay, and the observed pole–zero interlacing confirms the collocated actuator–sensor configuration in each axis.

%\newpage

\begin{comment}
    \begin{figure*}[t!]
	\centering
	\subfloat[$G_{xx}(s)$]{\includegraphics[width = 0.495\textwidth]{Figures/OL_G11_mag.png}}
	 \hfill
	\subfloat[$G_{xy}(s)$]{\includegraphics[width = 0.495\textwidth]{Figures/OL_G12_mag.png}}
    %\hfill
    
    \subfloat[$G_{yx}(s)$]{\includegraphics[width = 0.495\textwidth]{Figures/OL_G21_mag.png}}
	\hfill
	\subfloat[$G_{yy}(s)$]{\includegraphics[width = 0.495\textwidth]{Figures/OL_G22_mag.png}}
    %\hfill
    
        \caption{Identified frequency response plots of MIMO nanopositioning system $\boldsymbol{G}(s):=\boldsymbol{f}(s) \mapsto \boldsymbol{y}(s)$ (\protect\blackline).}
        \label{fig:SystemIdentification}
\end{figure*}
\end{comment}

\section{CONTROL Design Methodology}
\label{sec:DesignMethodology}
\begin{figure}[t!]
    \centering
    \includegraphics[width =0.9\linewidth]{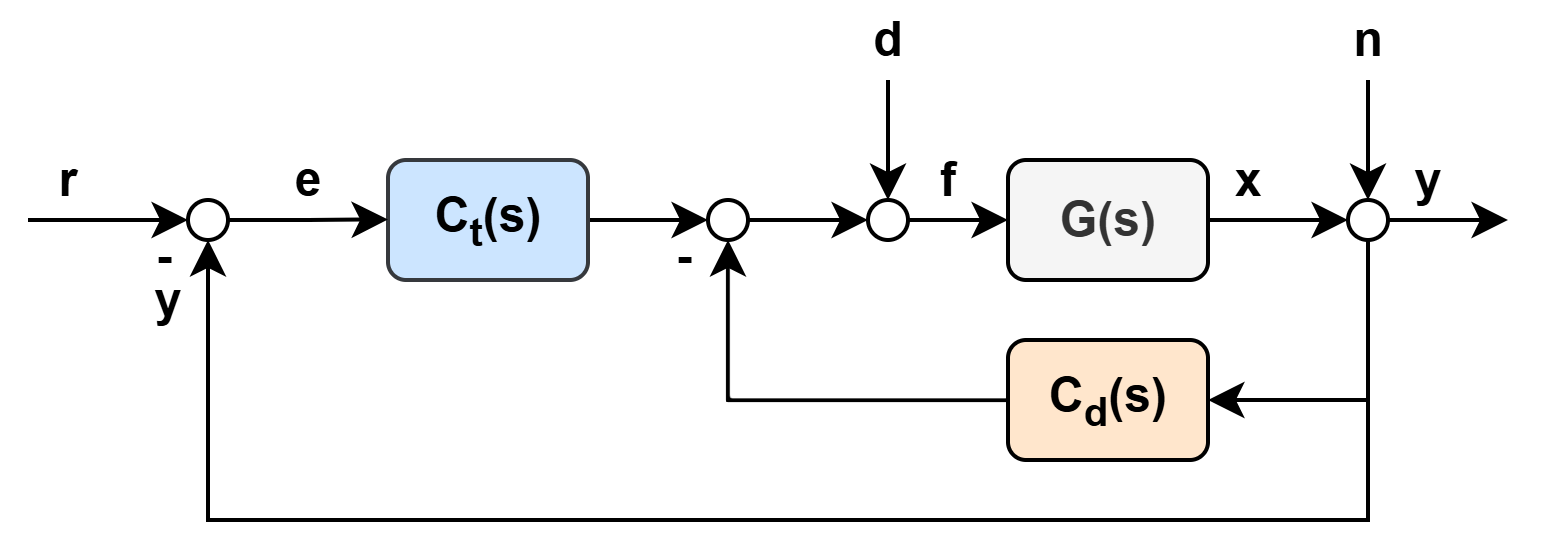}
    \caption{Dual closed-loop architecture incorporating damping controller $\boldsymbol{C_d}(s)$ and tracking controller $\boldsymbol{C_t}(s)$.}
    \label{fig:ControlArchitecture}
\end{figure}

This section outlines the control design methodology, following the dual-loop control architecture shown in Fig.~\ref{fig:ControlArchitecture}.

\subsection{Decentralized Control Architecture}
\label{sec:DecentralizedControlArchitecture}
The 2-axis nanopositioning system is modeled as a $2\times2$ MIMO transfer function matrix $\boldsymbol{G}(s)$:
\begin{equation}
\boldsymbol{G}(s)=
\begin{bmatrix}
G_{xx}(s) & G_{xy}(s) \\
G_{yx}(s) & G_{yy}(s)
\end{bmatrix},
\end{equation}
where $\boldsymbol{G}(s)$ maps the actuation input vector $\boldsymbol{f}(s)=\begin{bmatrix} f_x & f_y \end{bmatrix}^T$ to the sensor output vector $\boldsymbol{y}(s)=\begin{bmatrix} y_x & y_y \end{bmatrix}^T$ along the two axes. The vector $\boldsymbol{r}(s)=\begin{bmatrix} r_x \ r_y \end{bmatrix}$ denotes the reference inputs for each axis, while $\boldsymbol{d}(s)$ and $\boldsymbol{n}(s)$ represent the input and output disturbances, respectively.

The dual-loop architecture in Fig.~\ref{fig:ControlArchitecture} includes an active damping controller matrix $\boldsymbol{C_d}(s)$ in the inner loop and a tracking controller matrix $\boldsymbol{C_t}(s)$ in the outer loop. The inner-loop damping controller improves disturbance and noise rejection by suppressing the dominant resonance peaks of the system, thereby enabling the outer-loop tracking controller to be designed for higher bandwidth.

A decentralized framework is adopted to independently regulate each axis, resulting in diagonal controller matrices:
\begin{equation}
\boldsymbol{C}_d(s)=\left[\begin{array}{cc}
C_{d_x}(s) & 0 \\
0 & C_{d_ y}(s)
\end{array}\right], \quad \boldsymbol{C}_t(s)=\left[\begin{array}{cc}
C_{t_ x}(s) & 0 \\
0 & C_{t_ y}(s)
\end{array}\right] .
\end{equation}

The combined diagonal controller $\boldsymbol{D}(s)$ is defined as
\begin{equation}
\boldsymbol{D}(s):=\boldsymbol{C}_d(s)+\boldsymbol{C}_t(s)=\left[\begin{array}{cc}
D_x(s) & 0 \\
0 & D_y(s)
\end{array}\right],
\end{equation}
where $D_x(s)=C_{d_ x}(s)+C_{t_ x}(s), D_y(s)=C_{d_ y}(s)+C_{t_ y}(s)$.

\subsection{Damping and Tracking Control Design}
\label{sec:ControllerDesign}
In the identified system, each axis exhibits one dominant resonance peak, while the second resonance appears more prominently in the cross-coupling channels. To highlight cross-coupling reduction within a decentralized damping framework, two cases are considered in this paper: in Case (A), a single damping controller is used per axis to suppress the dominant resonance, whereas in Case (B), two damping controllers are implemented in parallel for each axis, with the second controller specifically targeting the second resonance observed in the cross-coupling terms.

To suppress the first dominant resonance, a non-minimum-phase resonant controller (NRC) is employed, defined as
\begin{equation}
\label{Eq:NRC_Controller}
C_{\text{NRC}j}(s)
= k_j \left(\frac{s - \omega_{a_j}}{s + \omega_{a_j}}\right),
\end{equation}
where $j \in \{x,y\}$, $k_j$, and $\omega_{a_j}$ denote the controller gain and corner frequency, respectively. The controller parameters are selected as $k_j = \gamma \cdot |G_{jj}(s)|^{-1}$ and $\omega_{a_j} = n\cdot\omega_{n1}$, where $\omega_{n1}$ is the first resonant frequency in each axis. Due to its non-minimum-phase structure, the NRC provides gain–phase decoupling, enabling the effective suppression of the targeted resonance mode  (\cite{natu2024non}).

Furthermore, to suppress the second resonance peak in Case (B), a band-pass controller (BPC) is added at the second resonant frequency $\omega_{n2}$ in each axis. The BPC, constructed as a cascade of second-order high-pass and low-pass filters, is defined as:
\begin{equation}
\label{Eq:BPC_Controller}
    \resizebox{1\hsize}{!}{$C_{\text{BPC}_j}(s) = \underbrace{\left(\frac{g_j s^2}{s^2+ 2\zeta_{1_j}\omega_{c_j} s+\omega_{c_j}^2}\right)\cdot\left(\frac{1}{s^2+ 2\zeta_{2_j}\omega_{c_j} s+\omega_{c_j}^2}\right)}_{\text{Band-Pass Filter}}\cdot\underbrace{\left(\frac{s-\omega_{d_j}}{s+\omega_{d_j}}\right)}_{\text{NMP Filter}},$}
\end{equation}
where a non-minimum-phase (NMP) filter with a corner frequency $\omega_{d_j}$ is included for delay compensation and to ensure closed-loop stability (\cite{natu2025non}). The band-pass structure ensures minimal influence on the low- and high-frequency closed-loop dynamics, with the control action concentrated around the targeted resonance frequency (\cite{natu5947393narrow}).

Thus, in Case (A), the damping controller for each axis is given by $C_{d_j}^{(A)}(s)=C_{\text{NRC}_j}(s)$, whereas in Case (B), the two parallel damping controllers yield $C_{d_j}^{(B)}(s)=C_{\text{NRC}_j}(s)+C_{\text{BPC}_j}(s)$.

For motion control in the outer feedback loop, a conventional proportional–integral (PI) controller in series with a first-order low-pass filter is employed:
\begin{equation}
\label{Eq:MotionController}
C_{t_j}(s)=\underbrace{k_{p_j} \cdot\left(1+\frac{\omega_{i_j}}{s}\right)}_{\begin{array}{c}
\text { Proportional } \\
\text { Integral Term }
\end{array}}\cdot\underbrace{\left(\frac{\omega_{l_j}}{s+\omega_{l_j}}\right)}_{\begin{array}{c}
\text { Low-Pass } \\
\text { Filter }
\end{array}},
\end{equation}
where $k_{p_j}$ is the proportional gain, $\omega_{i_j}$ is the integrator frequency, and $\omega_{l_j}$ is the low-pass cutoff frequency.

\begin{figure*}[t!]
	\centering
	\subfloat[Input in x-axis to output in x-axis.]{\includegraphics[width = 0.495\textwidth]{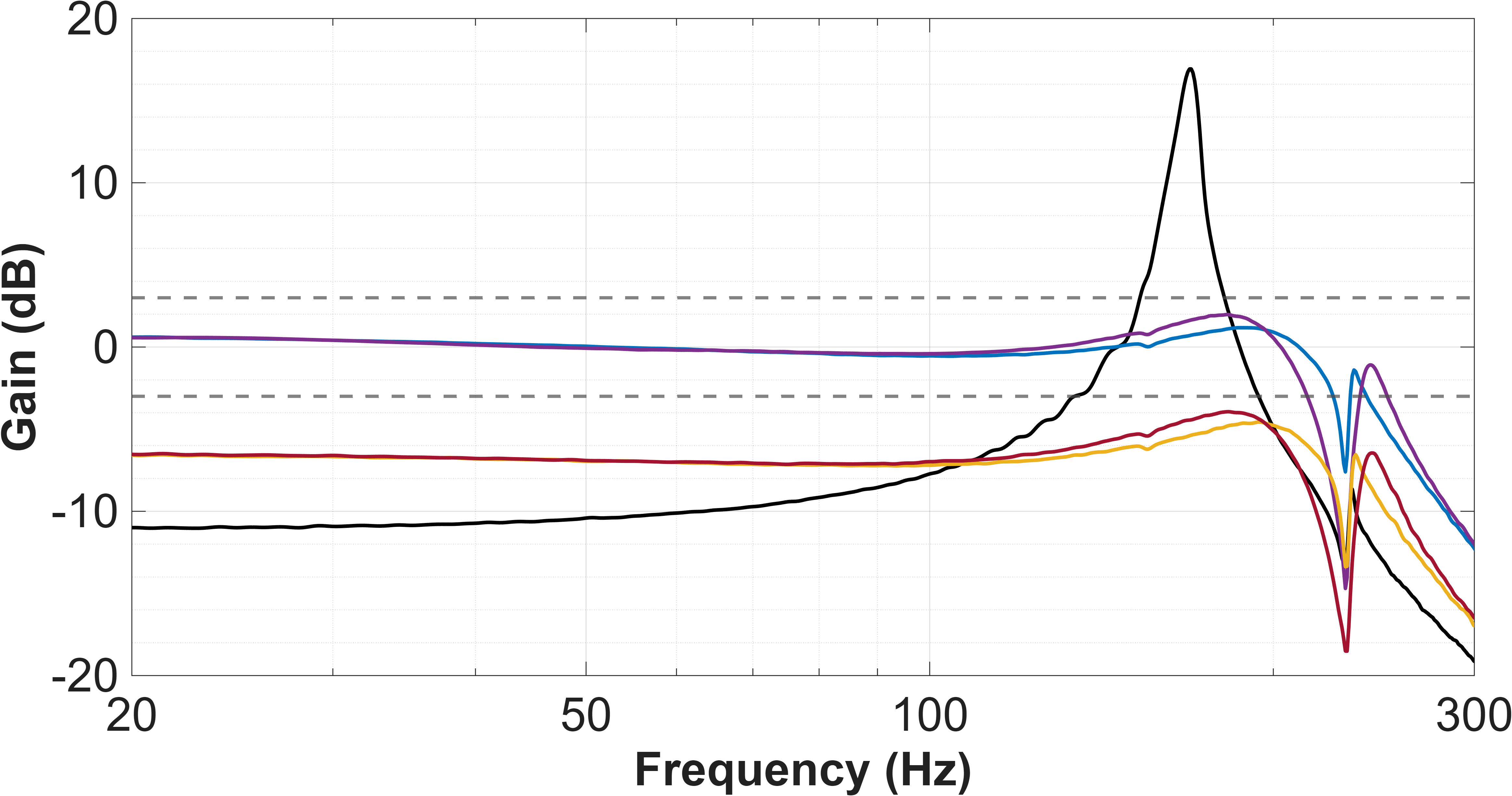}}
	 \hfill
	\subfloat[Input in y-axis to output in x-axis.]{\includegraphics[width = 0.495\textwidth]{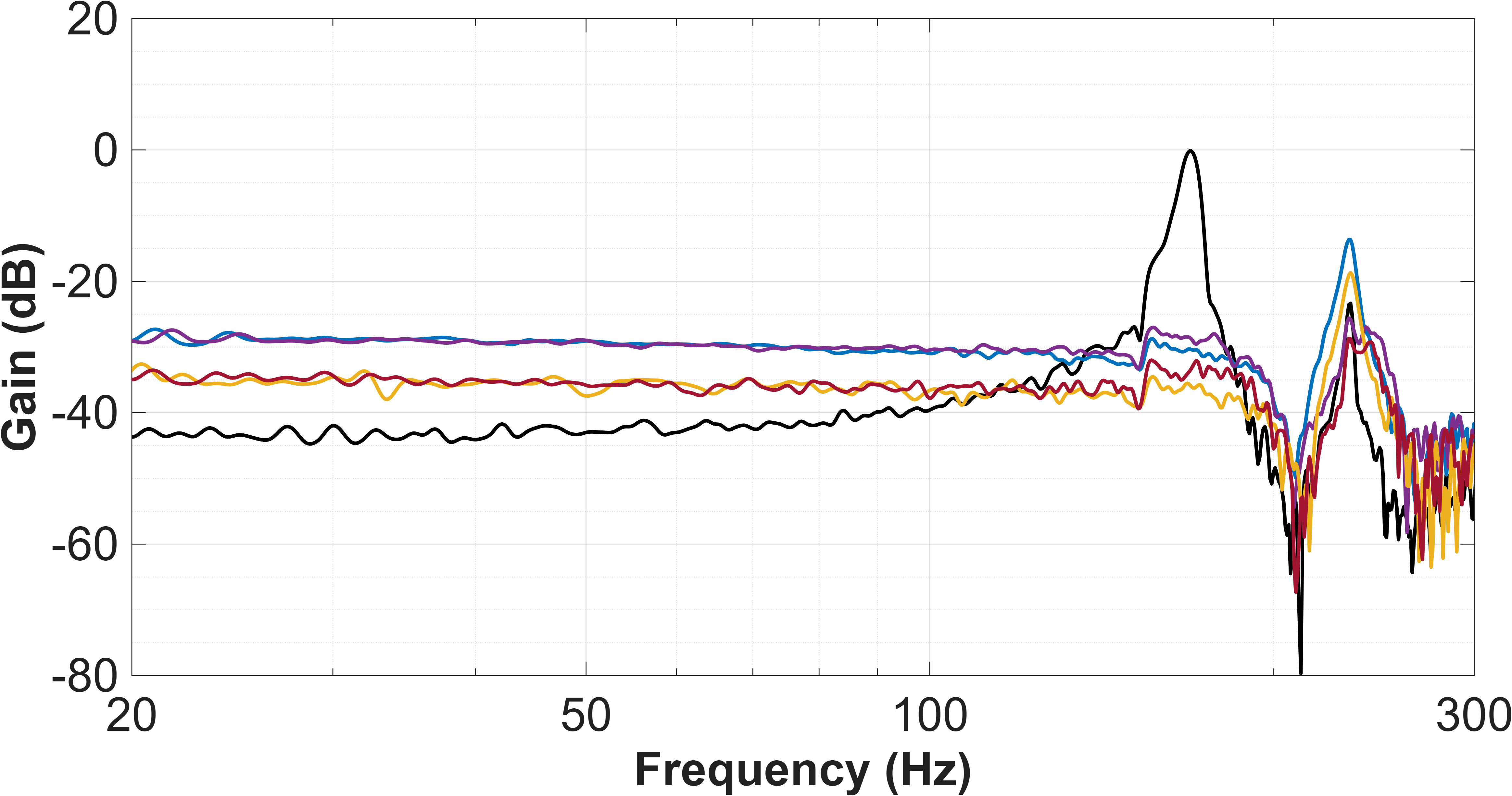}}
    %\hfill
    
    \subfloat[Input in x-axis to output in y-axis.]{\includegraphics[width = 0.495\textwidth]{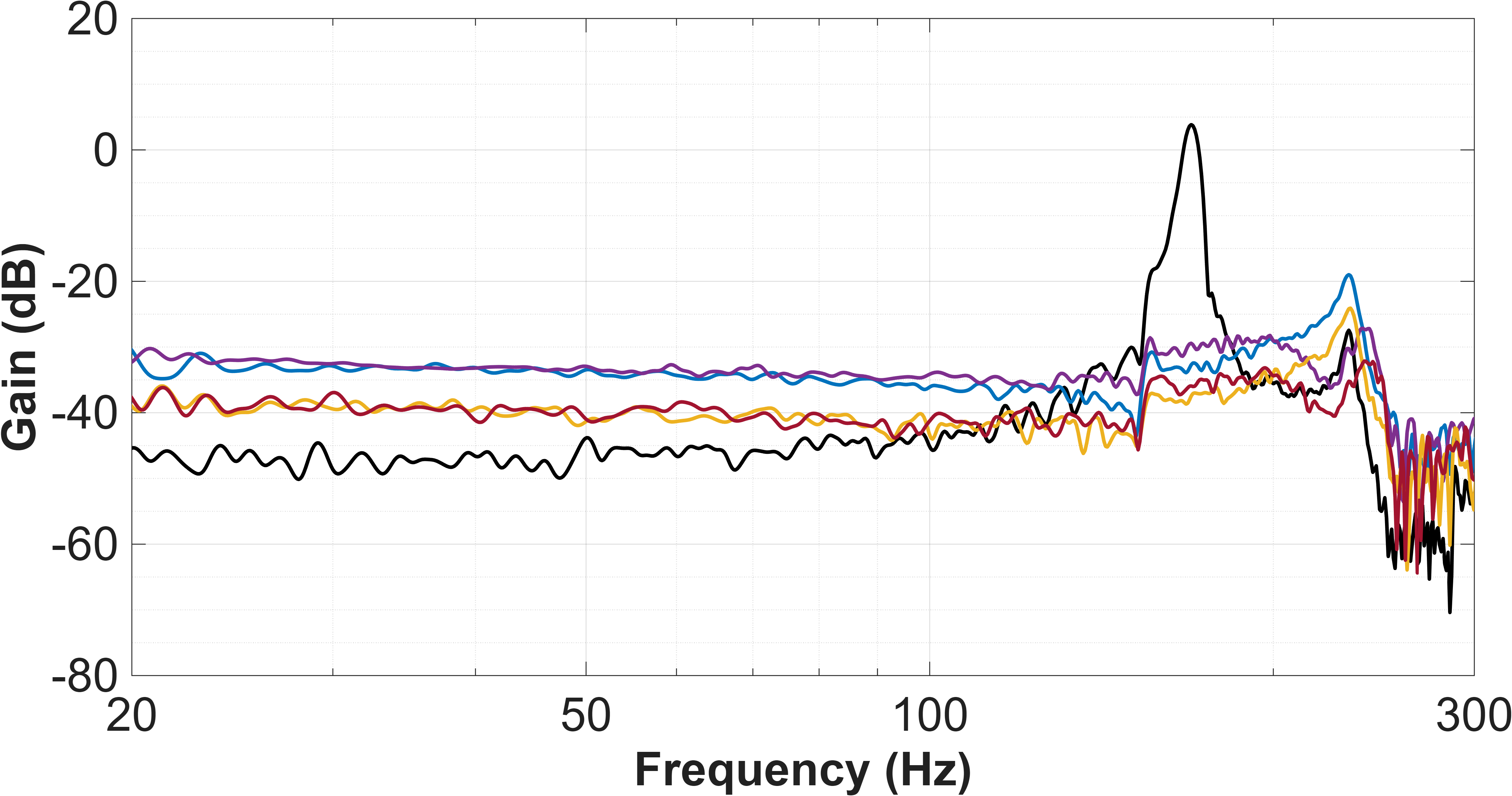}}
	\hfill
	\subfloat[Input in y-axis to output in y-axis.]{\includegraphics[width = 0.495\textwidth]{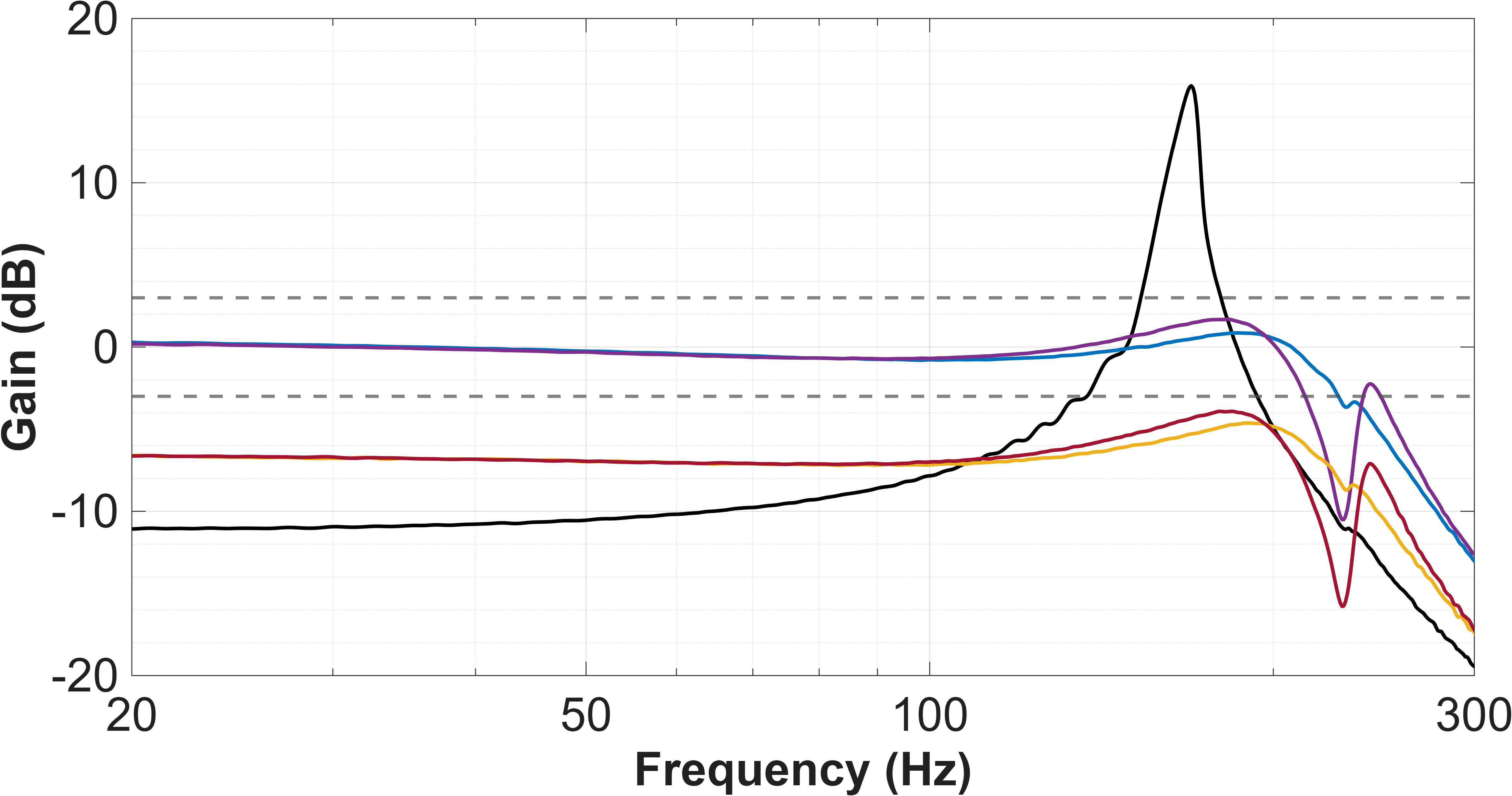}}
    %\hfill
    
        \caption{Experimental frequency response plots showing the identified MIMO system $\boldsymbol{G}(s)$ (\protect\blackline); closed-loop tracking sensitivities $\boldsymbol{T}^{(A)}(s)$ (\protect\blueline) and $\boldsymbol{T}^{(B)}(s)$ (\protect\purpleline) for Cases (A) and (B), respectively; and closed-loop process sensitivities $\boldsymbol{PS}^{(A)}(s)$ (\protect\yellowline) and $\boldsymbol{PS}^{(B)}(s)$ (\protect\maroonline) for Cases (A) and (B), respectively.}
        \label{fig:ClosedLoopFRF}
\end{figure*}

\subsection{Cross-Coupling Reduction}
\label{sec:CrossCouplingReduction}
With the band-pass controller in Case (B) operating in parallel within the inner damping loop, the objective is to reduce disturbance propagation between axes by suppressing the pronounced cross-coupling magnitude at the second resonance frequency in the closed loop. To evaluate this effect, the process sensitivity function $\boldsymbol{PS}(s)$ mapping disturbances $\boldsymbol{d}(s)$ to the measured output $\boldsymbol{y}(s)$ is examined, defined as:
\begin{equation}
\boldsymbol{PS}(s)=(\boldsymbol{I}+\boldsymbol{G}(s) \boldsymbol{D}(s))^{-1} \boldsymbol{G}(s),
\end{equation}
where $\boldsymbol{D}(s)=\operatorname{diag}(D_x(s), D_y(s))$. The matrix
$\boldsymbol{A}(s):=\boldsymbol{I}+\boldsymbol{G}(s)\boldsymbol{D}(s)$ then becomes
\begin{equation}
\boldsymbol{A}(s)=\left[\begin{array}{cc}
1+G_{x x}(s) D_x(s) & G_{x y}(s) D_y(s) \\
G_{y x}(s) D_x(s) & 1+G_{y y}(s) D_y(s)
\end{array}\right] .
\end{equation}

Let $\Delta(s):=\operatorname{det}(\boldsymbol{A}(s))$, then it can be computed to be:
\begin{equation}
\resizebox{1\hsize}{!}{$\Delta(s)=1+G_{x x}(s) D_x(s)+G_{y y}(s) D_y(s)+\operatorname{det}(\boldsymbol{G}(s)) D_x(s) D_y(s)$} ,
\end{equation}
where $\operatorname{det}(\boldsymbol{G}(s))=G_{x x}(s) G_{y y}(s)-G_{x y}(s) G_{y x}(s)$. The adjugate of $\boldsymbol{A}(s)$ is
\begin{equation}
\operatorname{adj}(\boldsymbol{A}(s))=\left[\begin{array}{cc}
1+G_{y y}(s) D_y(s) & -G_{x y}(s) D_y(s) \\
-G_{y x}(s) D_x(s) & 1+G_{x x}(s) D_x(s)
\end{array}\right] .
\end{equation}

Using $\boldsymbol{A}^{-1}(s)=\operatorname{adj}(\boldsymbol{A}(s)) / \Delta(s)$, we obtain;
\begin{equation}
\boldsymbol{PS}(s)=\boldsymbol{A}^{-1}(s) \boldsymbol{G}(s)=\frac{1}{\Delta(s)} \operatorname{adj}(\boldsymbol{A}(s)) \boldsymbol{G}(s)
\end{equation}
Thus, the process sensitivity function $\boldsymbol{PS}(s)$ becomes:
\begin{equation}
\resizebox{1\hsize}{!}{$\boldsymbol{PS}(s)=\frac{1}{\Delta(s)}\left[\begin{array}{cc}
G_{x x}(s)+\operatorname{det}(\boldsymbol{G}(s)) D_y(s) & G_{x y}(s) \\
G_{y x}(s) & G_{y y}(s)+\operatorname{det}(\boldsymbol{G}(s)) D_x(s)
\end{array}\right]$} .
\end{equation}

In particular, the reduction in both cross-coupling elements of $\boldsymbol{PS}(s)$ for any frequency $\omega$ is:
\begin{equation}
\left|\frac{\boldsymbol{PS}_{x y}(\mathrm{j} \omega)}{G_{x y}(\mathrm{j} \omega)}\right|=\left|\frac{1}{\Delta(\mathrm{j} \omega)}\right| \quad \And \quad\left|\frac{\boldsymbol{PS}_{y x}(\mathrm{j} \omega)}{G_{y x}(\mathrm{j} \omega)}\right|=\left|\frac{1}{\Delta(\mathrm{j} \omega)}\right| ,
\end{equation}
which are governed entirely by a single complex scalar $\Delta(\mathrm{j} \omega)=\operatorname{det}(\boldsymbol{I}+\boldsymbol{G}(\mathrm{j} \omega) \boldsymbol{D}(\mathrm{j} \omega))$.

Similarly, the cross-coupling reduction in closed-loop sensitivity $\boldsymbol{T}(s)=\boldsymbol{PS}(s)\boldsymbol{C_t}(s)$ can be computed as:
\begin{equation}
\left|\frac{T_{x y}(\mathrm{j} \omega)}{G_{x y}(\mathrm{j} \omega)}\right|=\frac{\left|C_{t, y}(\mathrm{j} \omega)\right|}{\left|\Delta_{\mathrm{tot}}(\mathrm{j} \omega)\right|} \quad \And \quad\left|\frac{T_{y x}(\mathrm{j} \omega)}{G_{y x}(\mathrm{j} \omega)}\right|=\frac{\left|C_{t, x}(\mathrm{j} \omega)\right|}{\left|\Delta_{\mathrm{tot}}(\mathrm{j} \omega)\right|} .
\end{equation}

\begin{figure}[t!]
    \centering
    \includegraphics[width =0.81\linewidth]{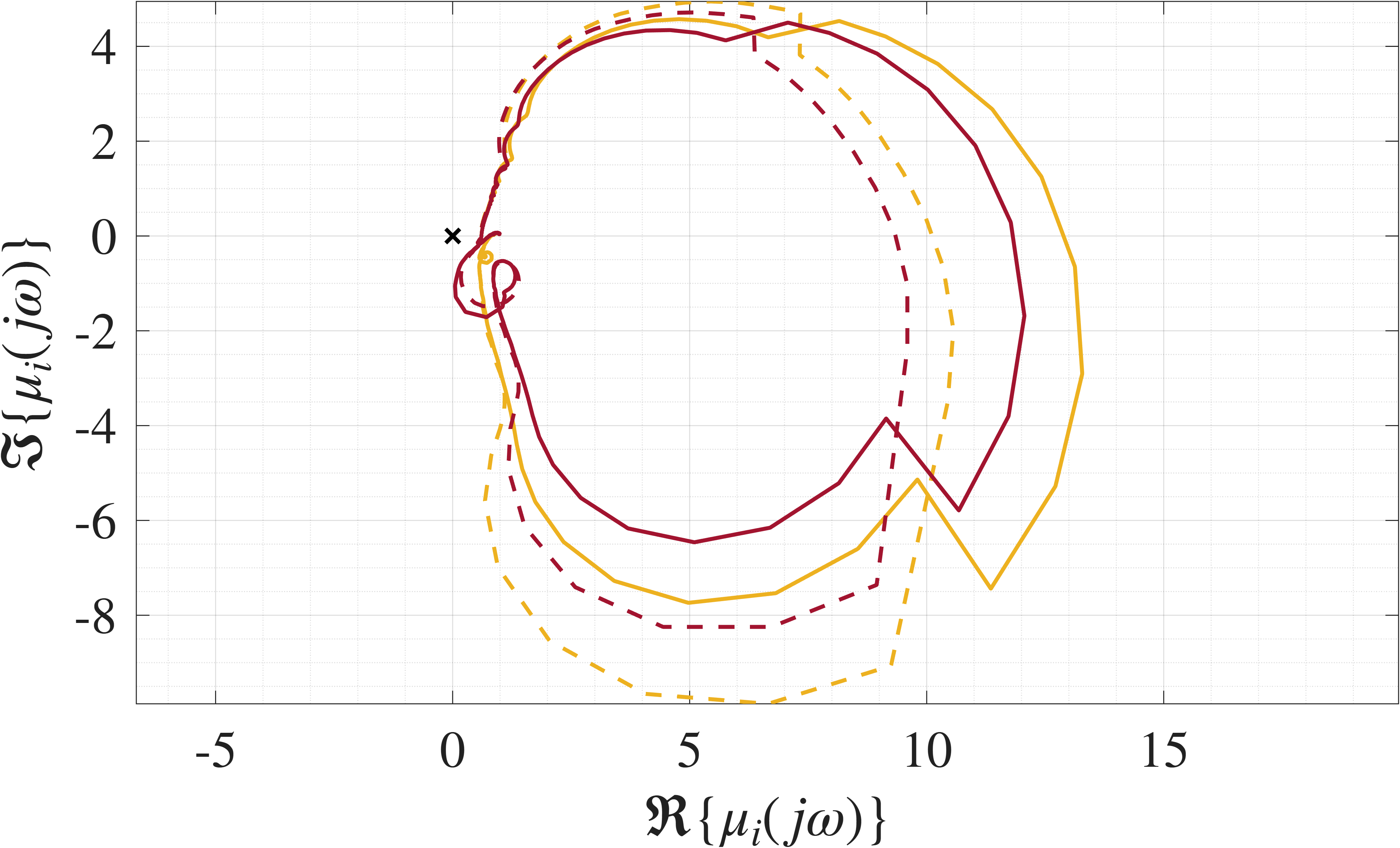}
    \caption{Nyquist loci of closed-loop eigenvalues for Case (A): $\mu_1(j\omega)$ (\protect\yellowline), $\mu_2(j\omega)$ (\protect\yellowlinedashed), and Case (B): $\mu_1(j\omega)$ (\protect\maroonline), $\mu_2(j\omega)$ (\protect\maroonlinedashed).}
    \label{fig:NyquistStability}
\end{figure}

\begin{figure}[t!]
    \centering
    \includegraphics[width =1\linewidth]{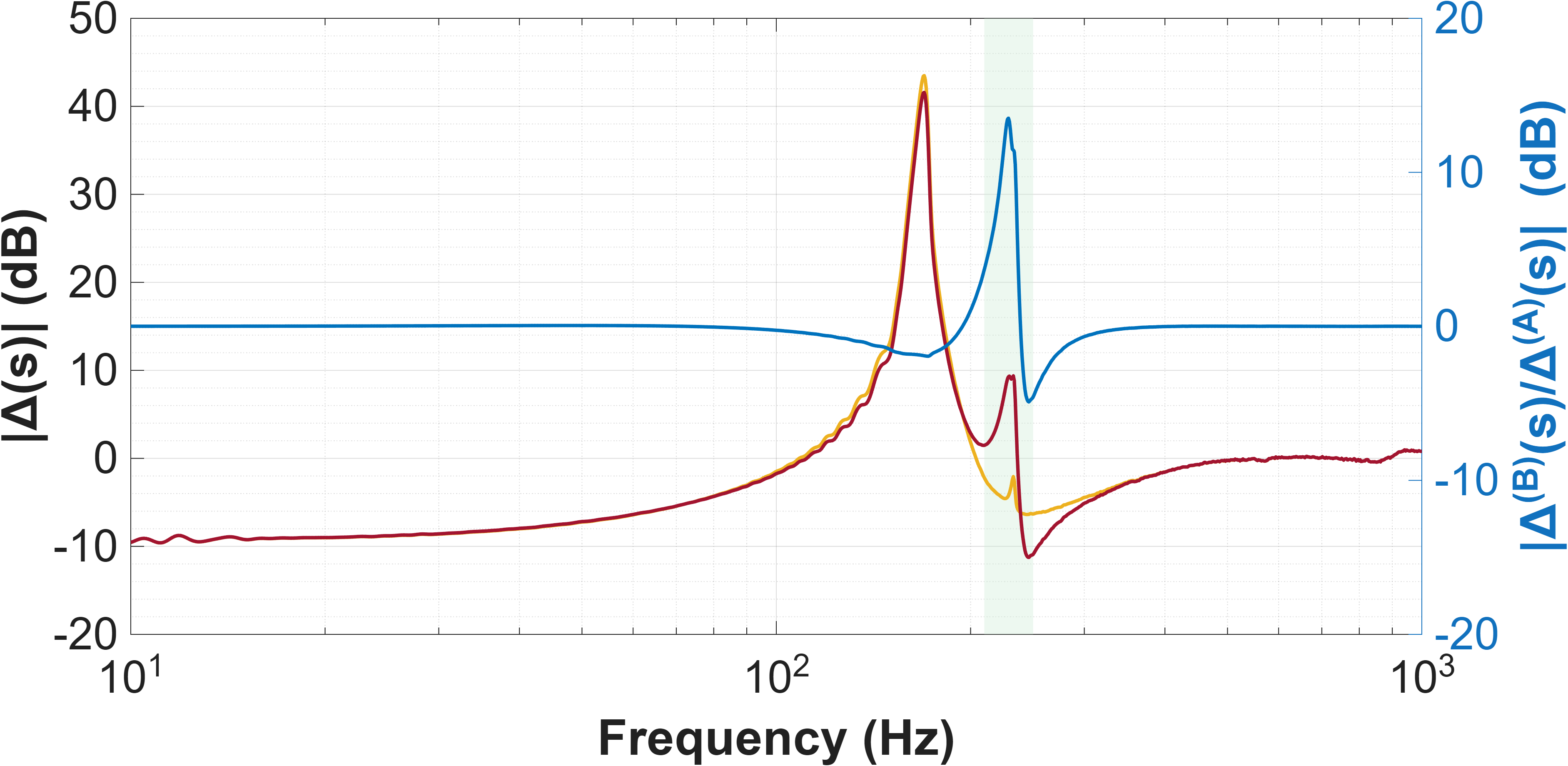}
    \caption{Magnitude of the cross-coupling reduction factor for Case (A), $|\Delta^{(A)}(s)|$ (\protect\yellowline), and Case (B), $|\Delta^{(B)}(s)|$ (\protect\maroonline), together with their ratio $|\Delta^{(A)}(s)|/|\Delta^{(B)}(s)|$ (\protect\blueline).}
    \label{fig:CrossCouplingReduction}
\end{figure} 

\section{Experimental Results}
\label{sec:ExperimentalResults}
This section presents the experimental validation of the proposed design methodology, with closed-loop performance demonstrated in both the frequency and time domains.
\begin{figure*}[t!]
	\centering
	\subfloat[]{\includegraphics[width = 0.495\textwidth]{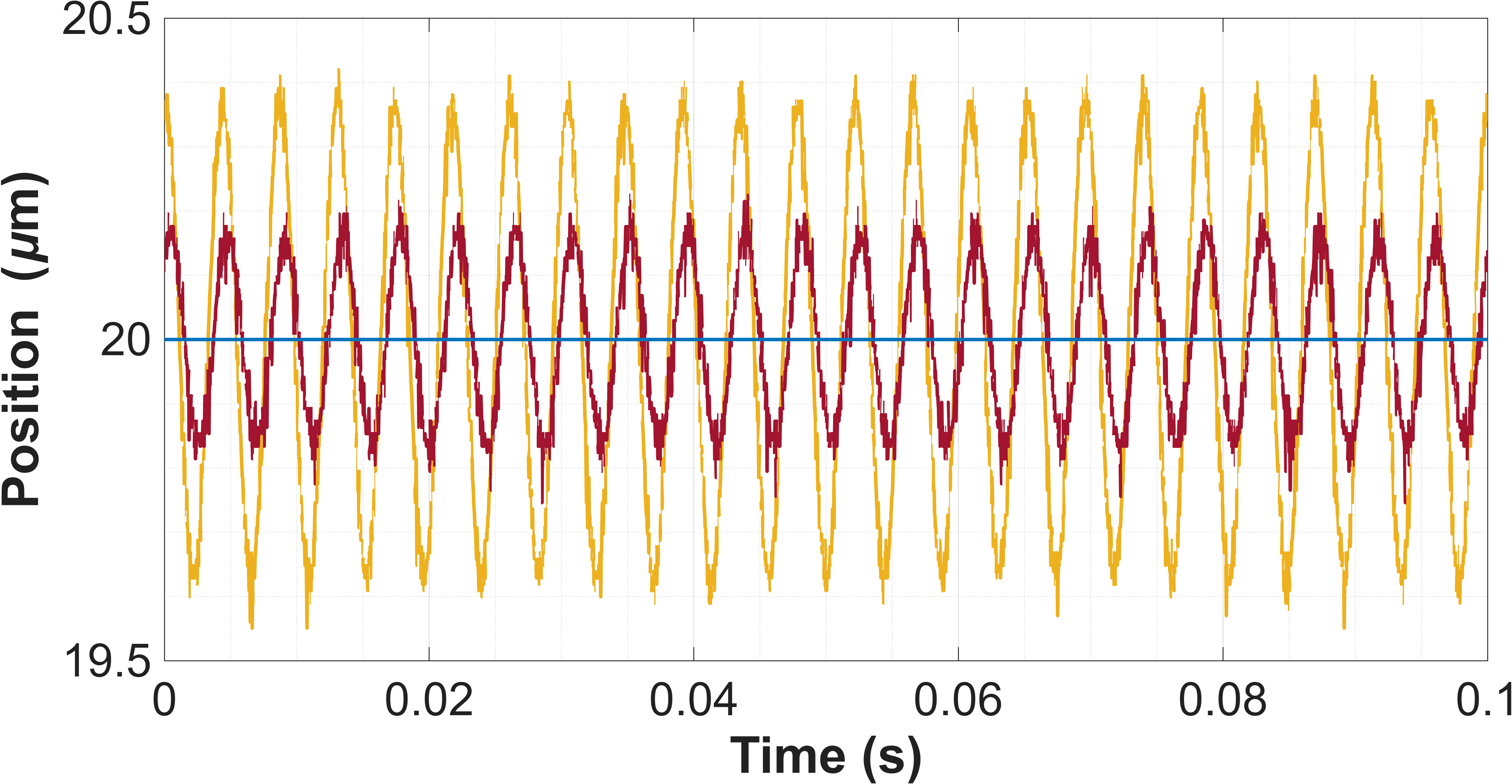}}
	 \hfill
	\subfloat[]{\includegraphics[width = 0.495\textwidth]{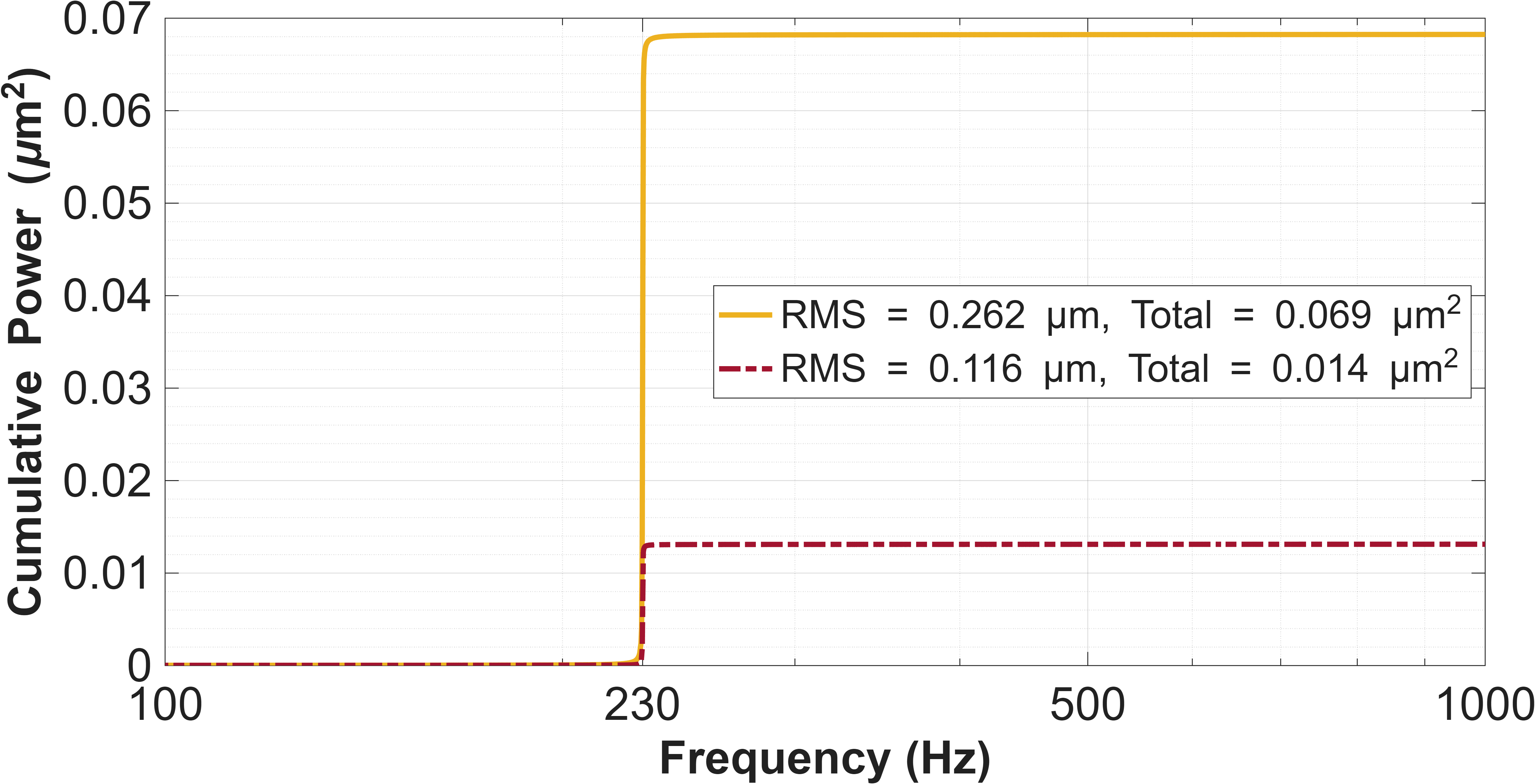}}
    %\hfill
        \caption{Comparison of disturbance rejection performance for Case (A) (\protect\yellowline) and Case (B) (\protect\maroonline): (a) measured Y-axis sensor output $y_y(t)$ and (b) its cumulative power spectrum, under a constant reference input (\protect\blueline) and a sinusoidal disturbance at 231~Hz.}
        \label{fig:DisturbanceRejection}
\end{figure*}
\begin{figure*}[t!]
	\centering
	\subfloat[]{\includegraphics[width = 0.495\textwidth]{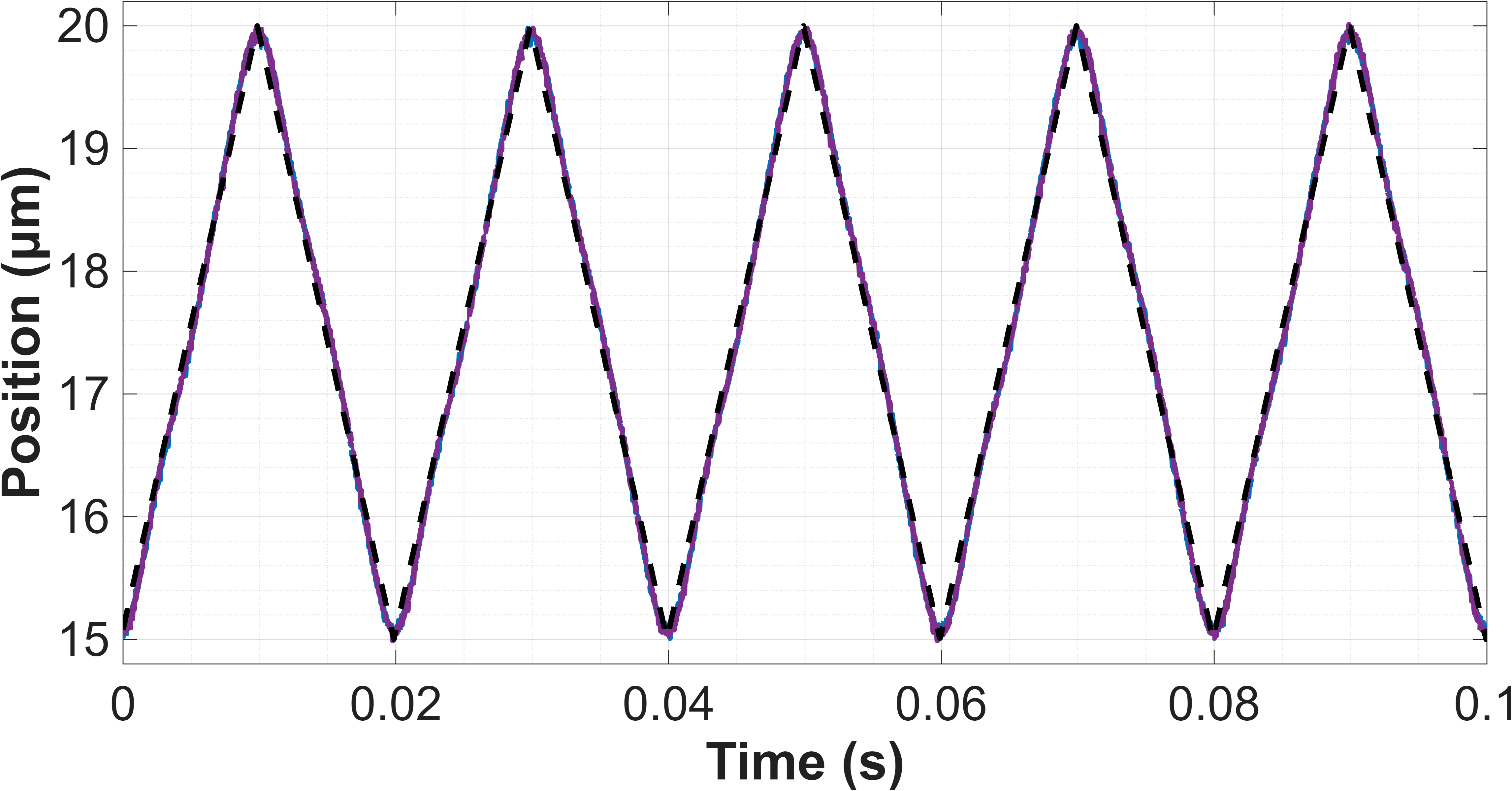}}
	 \hfill
	\subfloat[]{\includegraphics[width = 0.495\textwidth]{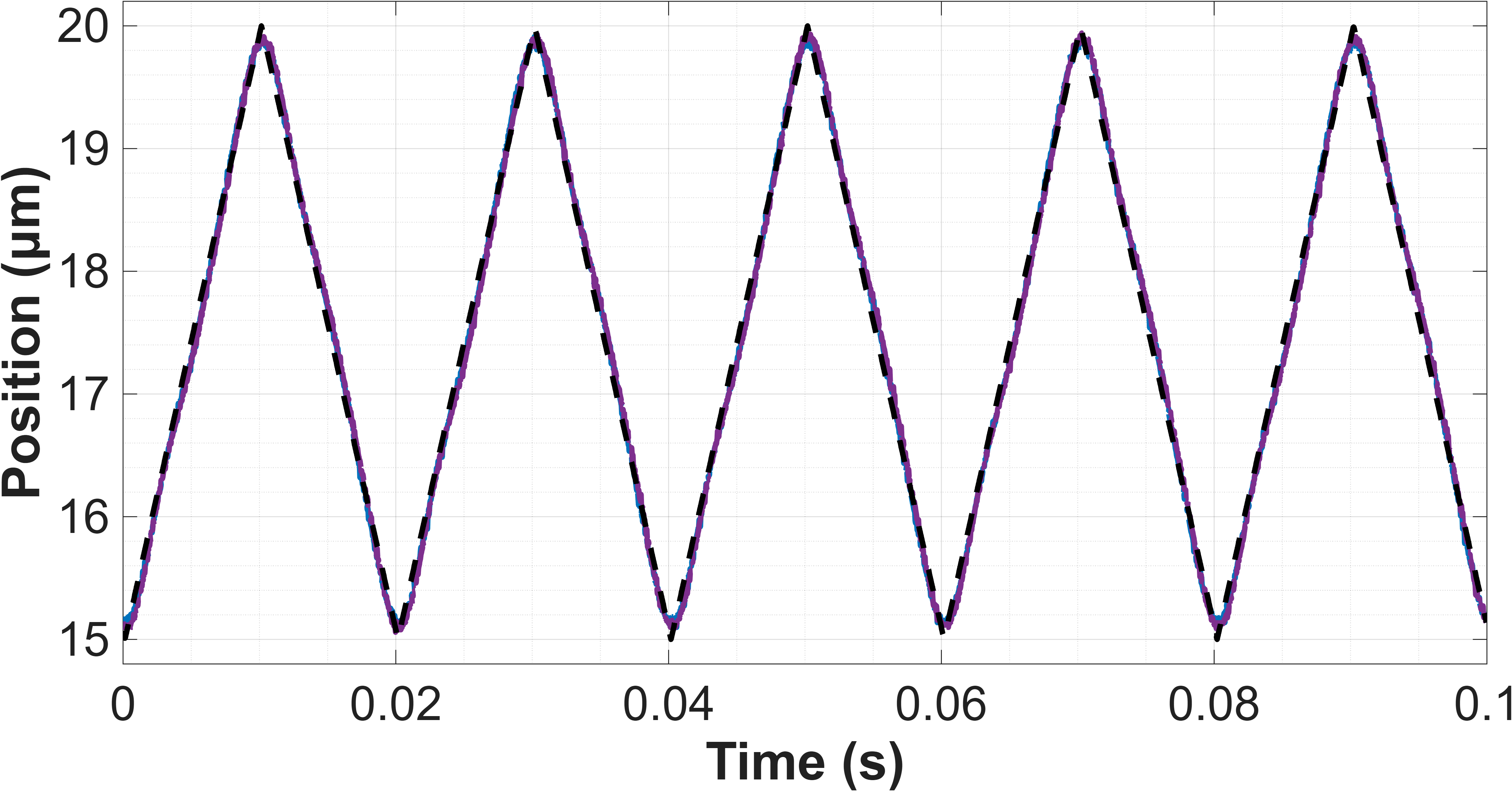}}
    %\hfill
        \caption{Comparison of reference-tracking performance for Case (A) (\protect\blueline) and Case (B) (\protect\purpleline) under a triangular reference input (\protect\blacklinedashed): (a) X-axis response with a constant reference on the Y-axis, and (b) Y-axis response with a constant reference on the X-axis.}
        \label{fig:ReferenceTracking}
\end{figure*}
\subsection{Closed-Loop Sensitivities and Cross-Coupling}
In accordance with the controller design presented in Section~\ref{sec:ControllerDesign}, the controller parameters are tuned to achieve maximum damping at the targeted resonances while enabling the largest possible $\pm 3$~dB closed-loop bandwidth for motion control. The resulting parameter values for each axis are summarized in Table~\ref{tab:ControllerParameters}.

With the tuned controllers, the stability of the MIMO closed-loop system is evaluated using a generalized Nyquist argument. For the loop transfer matrix $\boldsymbol{L}(s)=\boldsymbol{G}(s)\boldsymbol{D}(s)$, let $\lambda_i(s)$ denote its eigenvalues for $i=\{1,2\}$. The associated closed-loop characteristic matrix is $\boldsymbol{A}(s)=\boldsymbol{I}+\boldsymbol{L}(s)$, whose eigenvalues are then given by $\mu_i(s)=1+\lambda_i(s)$. The characteristic determinant of the MIMO feedback system is defined as $\Delta(s)=\operatorname{det}(\boldsymbol{A}(s))$. Using the standard determinant–eigenvalue relation, this can be written as:
\begin{equation}
\Delta(s)=\prod_{i=1}^2 \mu_i(s)=\prod_{i=1}^2\left(1+\lambda_i(s)\right) .
\end{equation}
For closed-loop stability, the Nyquist loci of the eigenvalues $\mu_i(j\omega)$ must not encircle the origin (accounting for any open-loop unstable poles), which is satisfied in both considered cases, as illustrated in Fig.~\ref{fig:NyquistStability}.

\begin{table}[t!]
    \centering
    \caption{Controller parameters tuned for experimental implementation.}
    \begin{tabular}{l|l|l}
\hline \textbf{Parameter} & \textbf{X-Axis} & \textbf{Y-Axis} \\
\hline $k$ ($\gamma$) & 3.5978 (0.995) & 3.5632 (0.995) \\
 $\omega_a$ ($n$) & 495~Hz (3) & 495~Hz (3) \\
 $\omega_c$ & 231~Hz & 231~Hz \\
 $g$ & 6.6567e5 & 6.6567e5 \\
 $\zeta_1$, $\zeta_2$ & 0.02, 1 & 0.02, 1 \\
 $\omega_d$ & 2000~Hz & 2000~Hz \\
 $k_p$ & 2.1466 & 2.0967 \\
 $\omega_i$ & 3.35~Hz & 3.35~Hz \\
 $\omega_l$ & 400~Hz & 400~Hz \\
\hline
\end{tabular}
    \label{tab:ControllerParameters}
\end{table}

Closed-loop performance is evaluated in the frequency domain using two experimentally identified sensitivity functions: the tracking sensitivity $\boldsymbol{T}(s):\boldsymbol{r}(s)\mapsto\boldsymbol{y}(s)$ and the process sensitivity $\boldsymbol{PS}(s):\boldsymbol{d}(s)\mapsto\boldsymbol{y}(s)$, as shown in Fig.~\ref{fig:ClosedLoopFRF}. The NRC implemented in both cases effectively suppresses the dominant first resonance in each axis, enabling high closed-loop bandwidths. In Case (A), the $\pm 3$~dB bandwidths are $225$~Hz and $228$~Hz for the x- and y-axes, respectively. In Case (B), the shaped frequency response of the combined damping controller yields bandwidths of $214$~Hz and $213$~Hz, still above the first resonant mode in each axis.

\begin{comment}
    Compared to the identified plant response, a magnitude reduction of $5.38$ dB and $7.34$ dB is achieved in $\boldsymbol{PS}_{xy}(s)$ and $\boldsymbol{PS}_{yx}(s)$, while relative to the closed-loop response using only one damping controller $\boldsymbol{C_{d}^{(A)}}(s)$, the reduction amounts to $10.05$ dB and $11.22$ dB in each channel, respectively. 
\end{comment}

However, the NRC introduces amplification of the second resonant mode in both sensitivity functions due to its constant-gain property. In Case (B), this effect is mitigated by the parallel damping structure $\boldsymbol{C_d^{(B)}}(s)$, where the added band-pass controller reduces the pronounced second resonance in the cross-coupling components of $\boldsymbol{PS}(s)$, as shown in Fig.~\ref{fig:ClosedLoopFRF}(b,c). This improvement is highlighted by examining the complex scalar $\Delta(s)$ over the frequency range of interest. Case (B) achieves an approximate $9.4$~dB reduction in the cross-coupling peak, whereas Case (A) exhibits a $2.1$~dB amplification. The ratio of the two cases, also plotted in Fig.~\ref{fig:CrossCouplingReduction}, demonstrates an overall improvement of approximately $11.5$~dB at the targeted second resonance when the band-pass controller is incorporated.

\subsection{Disturbance Rejection and Reference Tracking}
Disturbance rejection was evaluated using steady-state tracking experiments in which each axis followed a constant reference input while a sinusoidal external disturbance at the second resonance frequency (231 Hz) was injected into one axis. To demonstrate the effect of the band-pass controller on suppressing the cross-coupling resonance, both cases were tested under identical disturbance conditions. As shown in Fig.~\ref{fig:DisturbanceRejection}, the sensor-output fluctuations are reduced when the parallel damping controller is employed in the inner loop. This improvement is further illustrated in the power spectrum, where the cumulative power and corresponding RMS displacement indicate a significant reduction compared to the case without cross-coupling suppression. By attenuating the cross-coupling term in Case (B), the disturbance propagation between axes is substantially reduced, thereby enhancing the positioning precision for each axis.

The reference-tracking performance was assessed using a triangular input representative of a raster-scanning operation. A 50 Hz triangular reference ranging from 15 to 20~$\mu$m was applied to compare the two controller configurations. As expected from the similar closed-loop $\boldsymbol{T}(s)$ responses, both cases exhibit comparable tracking performance, with only minor differences in RMS tracking error along each axis. For the X-axis, the RMS error is $0.1167$~$\mu$m in Case (A) and $0.1127$~$\mu$m in Case (B), while for the Y-axis, the errors are $0.0964$~$\mu$m and $0.1124$~$\mu$m, respectively.

\section{Conclusion}
\label{sec:Conclusion}
A decentralized dual-loop control strategy was presented for a two-axis piezoelectric nanopositioner with lightly damped resonances and strong cross-axis coupling. An inner non-minimum-phase resonant damping controller, combined with an outer motion controller on each axis, enabled high-bandwidth tracking beyond the first structural mode. By augmenting the damping loop with a parallel band-pass path tuned to a higher-order resonance that predominantly affects the cross-coupling channels, the proposed design significantly reduces cross-axis coupling while preserving tracking accuracy. Frequency- and time-domain experiments confirmed improved damping, enhanced disturbance rejection, and increased cross-axis isolation, demonstrating the effectiveness of targeted band-pass damping in decentralized MIMO nanopositioning control.

\begin{ack}
The authors express their sincere gratitude to Mathias Winter, Head of Piezo System \& Drive Technology, and Dr.-Ing. Simon Kapelke, Head of Piezo Fundamental Technology, Physik Instrumente (PI) SE \& Co. KG, for their invaluable collaboration in providing technical insights concerning the system and its applications.
\end{ack}

\begin{comment}
    \section*{DECLARATION OF GENERATIVE AI AND AI-ASSISTED TECHNOLOGIES IN THE WRITING PROCESS}
During the preparation of this work, the authors used ChatGPT to refine the language and enhance readability. After using this tool, the authors reviewed and edited the content as needed and take full responsibility for the content of the publication. 
\end{comment}

\bibliography{ifacconf}             % bib file to produce the bibliography
                                                     % with bibtex (preferred)
                                                   
%\begin{thebibliography}{xx}  % you can also add the bibliography by hand

%\bibitem[Able(1956)]{Abl:56}
%B.C. Able.
%\newblock Nucleic acid content of microscope.
%\newblock \emph{Nature}, 135:\penalty0 7--9, 1956.

%\bibitem[Able et~al.(1954)Able, Tagg, and Rush]{AbTaRu:54}
%B.C. Able, R.A. Tagg, and M.~Rush.
%\newblock Enzyme-catalyzed cellular transanimations.
%\newblock In A.F. Round, editor, \emph{Advances in Enzymology}, volume~2, pages
%  125--247. Academic Press, New York, 3rd edition, 1954.

%\bibitem[Keohane(1958)]{Keo:58}
%\newblock \emph{Power and Interdependence: World Politics in Transitions}.
%\newblock Little, Brown \& Co., Boston, 1958.

%\bibitem[Powers(1985)]{Pow:85}
%\newblock Is there a way out?
%\newblock \emph{Harpers}, pages 35--47, June 1985.

%\bibitem[Soukhanov(1992)]{Heritage:92}
%\newblock \emph{{The American Heritage. Dictionary of the American Language}}.
%\newblock Houghton Mifflin Company, 1992.

%\end{thebibliography}

\appendix

                                                                         % in the appendices.

\end{document}